\documentclass[a4paper,aps,pre,showpacs,twocolumn,superscriptaddress]{revtex4}
\usepackage{amssymb}

\usepackage{graphicx}

\begin{document}

\title{Affinity driven social networks}
\author{B. Ruy\'u}
\affiliation{Centro At{\'o}mico Bariloche and Instituto Balseiro, 8400 S. C.
de Bariloche, Argentina}

\author{M. N. Kuperman}
\email{kuperman@cab.cnea.gov.ar}
\affiliation{Centro At{\'o}mico Bariloche and Instituto Balseiro, 8400 S. C.
de Bariloche, Argentina}

\begin{abstract}
In this work we present a model for evolving networks, where the
driven force is related to the social affinity between individuals
in a population. In the model, a set of individuals initially
arranged on a regular ordered network and thus linked with their
closest neighbors are allowed to rearrange their connections
according to a dynamics closely related to that of the stable
marriage problem. We show that the behavior of some topological
properties of the resulting networks follows a non trivial
pattern.
\end{abstract}
\maketitle

\vspace{1cm}

\section{Introduction}

The stable marriage problem, introduced by Gale and Shapley in
1962 \cite{gs}, is a well known example of an optimization
problem. In the original formulation,  men and women look for a
partner to marry. Previously, each agent, man or woman, ranks all
the individuals of the opposite sex according to a personal
preference. The only motivation that governs the evolution towards
a given configuration of marriages is to get married to someone at
the top of ones priorities list. But as the individual preference
is not necessarily symmetric,  the evolution towards a stable
situation is not trivial.

In general terms, the stable matching problem is a prototype model
in economics and social sciences,  where agents act following
selfish premises to optimize their own satisfaction, and with
underlying mutually conflicting constraints. However, the
emergence of global configurations promoted by more collective or
collaborative attitudes have also been studied. For example, in
\cite{niew} Nieuwenhuizen focussed on the properties of globally
optimal matchings which are advantageous for the society as a
whole, but not necessarily for all individuals.

It is not surprising  that besides its practical relevance, the
stable marriage problem presents many interesting theoretical
features that have attracted researchers from computer and social
sciences, mathematics, economics and  game theory
\cite{roth,roth1,roth2,gale1,df,ada}. The connection between the
stable marriage problem and classical disordered systems,
established in \cite{omer} promoted the interest on the problem
within the physics community \cite{niew,dzie,calda,ale}

Letting the system evolve from its initial configuration, we can
claim to have achieved a stable situation when it is not possible
to find a man and a woman not married each other that would prefer
to join themselves in  a new couple, leaving their corresponding
partners alone. Such a matching is called stable since no
individual has the chance to break it without remaining alone.
Previous studies have been mostly concerned with finding
algorithms for getting stable pairing configurations
\cite{GusIrv,knuth}. Given the rules according to which marriages
may be reconfigured, the main question  is whether a stable
situation where no benefits can be obtained from ulterior
rearrangement can be achieved. In \cite{gs} the authors proved
that each instance of the marriage problem has at least one stable
solution, and they presented an efficient algorithm to find it.

An instance of the stable matching problem is completely specified
by the preference or ranking matrix $X$, where all the information
about each individual preferences is contained. We can also define
the problems in social terms, by considering a marriage tension
associated to the mutual attraction. The more mutually affine the
individuals in a marriage are, the lower the marriage tension will
be. We can thus associate the cost of a marriage with the mutual
affinity. It is posible to store in the elements $x_{ij}$ and
$x_{ji}$ of $X$ the information about the tension that a marriage
between $i$ and $j$ will represent for each one of them
respectively. The energy or tension of a configuration of a
marriage can be defined as the sum of the individual tensions of
the each of the partners in the couple.

In the next sections we present the system under study and the
obtained results.

\section{The model}

\subsection{Social Affinity}
The system consists in $N$ interacting individuals represented by
the nodes of an evolving network. Though the number of nodes and
links will remain fixed throughout the whole process, networks can
evolve by changing their topology through the rewiring of the
links. The rewiring is done following the premise that each
individual will try to be connected to those individuals who are
socially more affine to him. This dynamics is different from the
dynamics of the stable marriage problem in that here individuals
try to conform groups not necessarily isolated instead of couples.
At this point we recall the concept of social affinity between a
couple of individuals. We understand that social affinity is the
result of the mutual interest that a given pair of individuals
arise in each other. The affinity is compose then by sum of the
personal interest or unilateral affinity of each of the
individuals in a given eventual relationship. This last feature
does not need to present symmetry properties. In an extreme
exercise of abstraction we quantify this ``emotional'' concept by
letting each individual to conform a list of her/his priorities at
the moment of choosing a partner or friend and assigning to each
of the other individuals a score ranging from zero to one. We
define thus the ranking matrix $X$, where the lower $x_{ij}$ is
the higher the interest that the node $i$ ``feels'' for the node
$j$ will be. As stated before, in general $x_{ij}\ne x_{ji} $, $
x_{ij} $ ranges from $0$ to $1$, and each node preference list is
made in a decreasing value of unilateral affinity. We can
establish an analogy between this system and a physical one, where
the tension plays the role of energy. Given a certain
configuration of the network, node $i$ is connected to, say
$n_{{\rm i}}$ other nodes, and we define first the energy of link
$ij$ as
\begin{equation}
e_{ij} = x_{ij}+x_{ji}
\end{equation}
The energy associated to the node $i$ is

\begin{equation}
E_{i} = \sum_{j\in \nu_i} x_{ij} \label{eneri}
\end{equation}
where $\nu_i$ denotes the neighborhood of $i$. Finally, we define
the global energy of the network, $E$ as follows
\begin{equation}
E = \sum_{i = 1}^{N} E_{i}= \sum_{i = 1}^{N-1} \sum_{\hat
j\in\nu_i}^{N} x_{i \hat j} \label{energ}
\end{equation}
where the sum is over the individuals $\hat j$ in $\nu_i$ such
that $\hat j > i$

The choice of the values that conform $X$ is of fundamental
importance in the development of the interactions, and therefore
in the outcome of the dynamics. Hence, special attention is paid
to this feature of the model resulting in two different
approaches, distinguished by the characteristics of the (initial)
distribution of  affinities. The first one, namely \textit{random
distributed}, assigns to the elements of $X$ random values between
$0$ and $1$, classified  into a certain number of discrete steps.
A social interpretation of this can be made in terms of a case
when people to whom one would like to be linked, follows no
special pattern within the network, and there is no correlation
between the mutual affinity and the physical or geographical
proximity. That means that (initial) vicinity or popularity plays
no role on the affinity development. This could be the case if the
system under study consist in individuals with no previous
knowledge of the others or if the index associated to the node is
for identification purposes only. As a mean to quantify the
initial range of variability of the affinities we allow the random
distributed variable to adopt only a  discrete set of values (or
steps). The extreme cases correspond to those when a) the
affinities can tale only two values and b) when the values are
uniformly distributed, with the number of steps only limited by
the discrete nature of our algorithm

\begin{figure}[hbt]
\centering \resizebox{\columnwidth}{!} {\rotatebox[origin=c]{0}{
\includegraphics{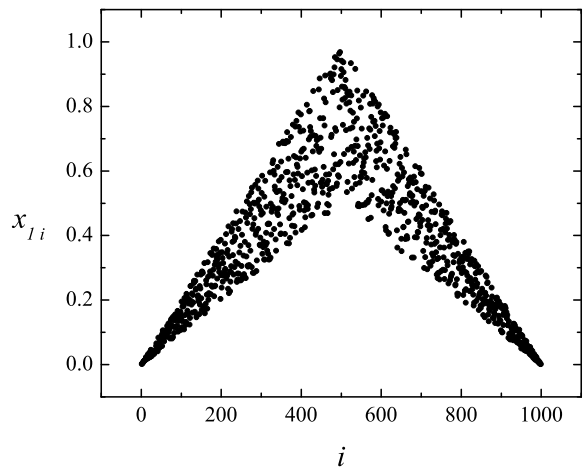}}}
\caption{Example of the tension values distribution of node 1,
when $N=1000$ and $q=0.5$} \label{afin}
\end{figure}

In the second approach, which will be referred as \textit{neighbor
correlated}, the affinity, on the contrary,  is closely related to
vicinity. Given a couple of nodes $i $ and $j$, we define the
normalized periodical distance between them, \textit{dist(i,j),}
as:
\begin{equation}
dist(i,j) = {\frac{{2}}{{N}}}{\left[ {{\left| {i - j}
\right|}\bmod {\frac{{N}}{{2}}}} \right]}.
\end{equation}
Then, we define $x_{ij} $ by the following expression:
\begin{equation}
x_{ij} = dist(i,j){\left[ {q + (1 - q)r} \right]}.
\end{equation}
where $r$ is a random number between $0$ and $1$, and $q$ a
parameter which we will call \textit{slope}. In an immediate
analysis of the above expression, it can be seen that the value of
tension (the inversely proportional to the  affinity) as a
function of normalized distance is a random number restricted to
the interior of two straight lines of slope $\pm 1$ and $\pm q$
respectively. In Fig. \ref{afin} we can see an example of the
distribution of values around node 1 in a system of 1000
individuals and $q=0.5$. The social correlate of this choice in
the distribution of affinity is to privilege closer neighbors when
choosing who to be connected with. One can imagine a number of
examples where this is the case and clearly, the vicinity or
spatial proximity affects the affinity between two individuals.

In this case, we want the spatial distribution to affect the
affinity  but not to ultimately determine it.  This would be the
case if $r$ was a constant number and  the distribution of
affinity would be so that first individuals in every list will be
the first neighbors and so on. Moreover, equally distant neighbors
would have equal affinity, due to the circular symmetry in the
definition of distance. One can infer, that as in the case of the
stable marriage problem, the dynamic of the system will be
governed by the affinity between individuals, in the sense that an
individual will try to be connected to others in the first
positions in her/his list. If we consider the case when $r$ is
constant, the final state of the system would necessary be a
regular network, in which each node is connected to her/his firsts
neighbors. To avoid this trivial case  we have introduced a random
distribution for $r$. Furthermore, the parameter $q$ will be
closely related to how close to the trivial case the final
configuration will be, reaching a complete matching when $q=1$. On
the other hand, choosing $q$ close to 0 will lead the system to be
closer to a random network. In other words, $q$ (or even better
$1/q$) can be considered as an order parameter of the network. All
these vague concepts regarding the final state of the system will
be formalized after defining the dynamics.

\subsection{Network Dynamics}

Once the affinities list has been assigned to each individual we
let the social configuration or network evolve. We start with
nodes conforming an initially ordered network and let the dynamics
associated to the traditional marriage problem lead the social
structure to a more stable (less energetic) configuration. The
governing dynamics may be chosen to minimize the individual
energy, Eq. (\ref{eneri}), or the global one, Eq. (\ref{energ}).
We will refer to these cases respectively as  (\textit{global
dynamics} and \textit{individual dynamics }). We will show that in
all the considered cases the system also achieves a configuration
with both lower energies.

It is important to point out that on this first stage we preserve
the number of initial links. The process of network rewiring
comprises two  aspects that must be clearly defined. On one side
there is an affinity based probability to break an already
established link to create a new one. On the other hand, we may
consider that each individual has a minimum number of associated
links, thus all the links will have an extreme attached
permanently to a given node. These are what we call one side-fixed
edges. The possibility of no a priori attached links will be also
considered as well.

\subsubsection{Case 1: One side-fixed edges}

We will start by considering the case when each node has $u$ edges
attached to it by one of the extremes, the connectivity of each
node is at least $u$, and the mean connectivity is $2u$. Mind that
this restriction does not imply that the edges are directed, since
once the evolution has finished, the edges are considered
undistinguishable. The further rewiring of a given link can be
chosen in terms of a global or individual energy minimization
rule. Thus we distinguish between the following two cases

a) Global decision dynamics (GDD)

In this case, the changes in the configuration of the network are
oriented to minimize the global energy $E$. Since the energy
landscape can be extremely complicated most of the times, the
system will get trapped in a configuration associated with a local
minimum. The mean individual energies will still be  quite high,
representing a discomfort situation among the individuals. To
avoid this situation we recur to a scheme based on simulated
annealing \cite{kirk,met}, therefore allowed changes will have a
non-zero probability of increasing the global energy, which in the
end will help to achieve lower value of $E$.  To continue the
analogy with the real annealing, we define a \textit{temperature
T} , which will be responsible for the fluctuations in the global
energy. The process starts with an ordered network at a certain
initial $T$. At each time step, a node and one of its links are
randomly chosen. The evaluation of the rewiring of the chosen
links follows, by measuring the change in energy involved in this
procedure, namely $\Delta E$. The change is accepted with
probability $p_a$ with
\begin{equation}
p_a= \left\{ \begin{array}{ccc} 1&\mbox{if}&\Delta E<0\\
\frac{1}{2} \exp({\frac{-\Delta E}{T}})&\mbox{if}&\Delta E> 0
\end{array}
\right.
\end{equation}

After a certain amount of time steps  (each one comprising $N$
computational steps), the threshold value $T$ is reduced, and we
iterate until the chosen minimum temperature is achieved. This
temperature usually is chosen to let the system reach a steady
state. Since changes in the network configuration are made taken
into consideration the global benefit, we can say that the
individuals are not ``selfish'' or that they are not in charge of
the evolution. If, for example, we break the link $(i,j)$ and
create the link $(i,k)$ the changes in global energy and in $i$'s
energy are, respectively
\begin{eqnarray}
\Delta E &=& \frac{1}{2}\left( {x_{i k} + x_{k i} - x_{i j}- x_{j
i}} \right)_{} \\ \nonumber \Delta E_{i} &=& x_{i k} - x_{i j}.
\end{eqnarray}
It is no difficult to imagine a situation in which $\Delta E$< 0
and $\Delta E_{{\rm i}{\rm} }$> 0. In this case, the change is
accepted, even though from $i$'s point of view is not convenient.

b) Individual decision dynamics 1 (IDD1)

In this case we consider that the individuals make selfish
decisions, since only the individual benefit is taken into account
at each change.  The mechanism here is very much the same as the
previous case, the only difference is that instead of using the
value $\Delta E$ as a threshold we take $\Delta E_{{\rm i}{\rm}
}$. Thus, we can not call the process a simulated annealing
anymore though the acceptance of changes is made according to the
probability $p_b$ defined as

\begin{equation}
p_b= \left\{ \begin{array}{ccc}
1&\mbox{if}&\Delta E_i<0 \\
\frac{1}{2}\exp({\frac{-\Delta E_i}{T}})&\mbox{if}&\Delta E_{i}>0
\end{array}
\right. \label{probb}
\end{equation}

Even though changes are made in a way to benefit the individual
who is making the decision, without regarding what happens with
the global energy, it can be seen that $E$ tends to decrease in
this process, as shown if Fig. \ref{energias} .

\subsubsection{Case 2: Free edges}

c) Individual decision dynamics 2 (IDD2)

The reason for this different approach is that we want deal with a
more realistic individual oriented dynamic. This time, edges are
no longer attached to any node. The only restrain in the dynamics
is to preserve the number of link and the connectivity of the
underlying network. Therefore we impose that no node is left
without a link, in other words, we do not accept isolated
individuals. Though this case is similar to IDD1 with $u=1$ tehre
are some important differences. Again, we proceed within a
simulated annealing frame, but the way we reconnect the network in
every time step changes.

The new mechanism goes as follows: we choose a node $i$ randomly,
and look at every node $i$ is connected to, choosing the one with
lower affinity (higher energy value), lets say $j$. Now we choose
another node $k$ at random not connected to $i$, and accept to
link $i$ with $k$ (replacing $j$) with a probability given by Eq.
(\ref{probb}). The main difference between this dynamics and the
previous one, IDD1, is that here we don't assign $u$ edges to
every node, so when it comes to decide which link to cut, the
choice does not restrict to what we previously called {\it
assigned edges}, but to any edge the node has. Furthermore, as the
decision whether to accept a connection or not is taken by only
one of the pair of linked nodes, individuals now have the
possibility to detach from those undesirable connections. This
balance plays a very important role in self organization.

\section{Numerical Results}
In what follows we will describe the results corresponding to the
cases of GDD, IDD1 and IDD2, obtained after extensive numerical
simulations of the described model.

Most of the simulations were done with networks of 100
individuals, with mean connectivity equal to 4 ( $u = 2$). When
effects that could be associated to size effects appeared, we
increased the size of the system to confirm our results.

\begin{figure}[hbt]
\centering \resizebox{\columnwidth}{!} {\rotatebox[origin=c]{0}{
\includegraphics{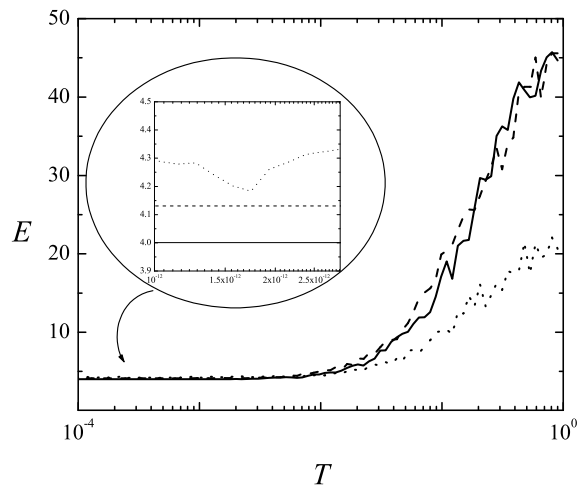}}}
\caption{Global energies as a function of temperature for the
cases GDD (full), IDD1 (dashed), and IDD2(dotted), with $q=0.35$
and $N=1000$. In the inset we show a detail of the evolution for
larger times i.e. smaller T} \label{energias}
\end{figure}

As said, for GDD we expected the system to achieve a steady state
(hopefully the minimum energy state), where no more changes take
place. Meanwhile in IDD1 and IDD2, we thought of the system self
organizing in a state of stationary global energy, where changes
still take place. However, both for random and neighbor correlated
distributions of affinity, we found  that GDD and IDD1 achieved
steady states, with similar values of the final global energy
(GDD's value is slightly smaller). On the other hand, IDD2
reaches, as expected, a stationary situation just as the one
described above, see Fig. \ref{energias}.  The fact that GDD
reaches a steady state is a direct result of how the dynamics is
defined. Once the system reaches an energy minimum and the
temperature is not high enough to displace it from there, no more
changes can take place. The constraint imposed on one of the
extremes of each link also drives the system to a steady situation
in the case of IDD1.

The absence of this constraints is what lets the system evolving
under the IDD2 scheme not to freeze in a given configuration. Once
in a configuration with a minimum energy, the system continues to
explore all the available changes leading to other configurations
without a change in the energy. It is now completely possible for
a change to take place in a way that global energy increases, but
the individual energy (of a certain node) decreases, being the
latter the one governing the dynamics.

All the three final states present close values of global energy,
generally differing in less than $10 \%$(see insent in Fig.
\ref{energias}), indicating that though the dynamics may be
associated to global or individual decisions, the network reaches
an ordered state.

According to the way how matrix $X$ was defined, it is apparent
that, except in a number of cases that depends on $q$, individuals
will prefer to be connected to their closer neighbors. We wonder
what kind of network topology  will be the final output of a
dynamics dominated by this feature.

\begin{figure}[hbt]
\centering \resizebox{\columnwidth}{!} {\rotatebox[origin=c]{0}{
\includegraphics{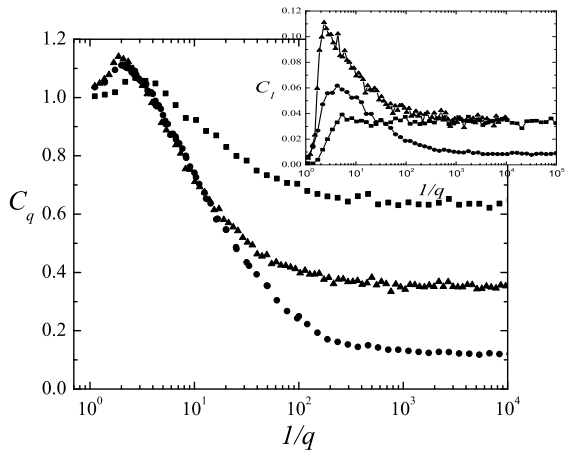}}}
\caption{Normalized clusterization as a function of $1/q$, for the
cases GDD (squares), IDD1 (circles), IDD2 (triangles). In the
inset we show the density of nodes $c_1$ for the same cases. With
$N=100$, $u=2$} \label{clust}
\end{figure}

We are interested in knowing whether the dynamics of the process
can lead to a structure with enhanced clusterization or not,
stressing even more the local organization of an ordered network.
We study thus, the clustering features of the resulting networks,
normalizing the obtained values to the clustering coefficient of
an ordered network. Our first observation is that when considering
a randomly distributed affinity the final configuration presents a
normalized clustering coefficient $C_q$ considerable lower than 1,
indicating that no apparent local structures are being formed. On
the contrary,  when considering an affinity correlatively
distributed, the clustering presents a marked non monotonous
dependence with parameter $q$. In figure \ref{clust} we can
observe that for large $1/q$ values (i.e. disordered distribution
of affinity) the normalized clustering coefficient is low, in
accordance to what can be expected in a disordered network.
However, when $1/q$ gets closer to zero, an anomalous behavior
takes place, the normalized clustering increases to values greater
than $1$, meaning that the network clustering is higher than the
corresponding value  for an ordered network with equal number of
nodes and links. Since clustering is a measure of connectivity
among close neighbors, the fact that clustering is greater than
the reference value reflects that some kind of closed structures
are being formed.  Given the correlated distribution, individuals
tend to favor connections to closed neighbors, but the interplay
with a certain degree of disorder introduced by $q$ results,
within a certain range of values of $q$, in a more complex final
configuration, with the emergence of a more local structure. It is
worth mentioning that there is a peak in the clustering, meaning
that there is a certain value $q$ presenting optimal  clustering.
This is a non trivial result that can affect the transport
properties of the network at local levels \cite{effi}.

Another quantity that reflects this behavior is what we denominate
$C_1$ nodes, defined as the number of nodes $c_1$ with absolute
clustering equal to $1$. A node being $c_1$ means that all its
neighbors are connected between them. A high mean clustering
coefficient might be related with a high number of $c_1$ nodes in
the network. When analyzing the behavior of $C_1$ as a function of
$q$ (Inset of Fig. \ref{clust}) we find that it is similar to the
one displayed by the clustering. It reaches a maximum at a finite
value of $q$. This reinforces the idea that the networks evolves
to a locally compact configuration, with nodes tending to conform
closed groups. When we say closed groups we do not mean
disconnected subgraphs but a nodes  highly connected among them
defined a cluster connected to the rest of the networks through a
few number of links.

\begin{figure}[hbt]
\centering \resizebox{\columnwidth}{!} {\rotatebox[origin=c]{0}{
\includegraphics{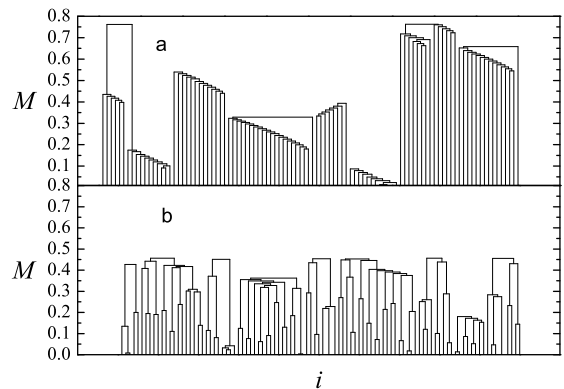}}}
\caption{Examples of the dendograms of the resulting networks for
IDD2,  $N=100$ and a) $q=0.48$,  b)$q=10^{-4}$ } \label{dendo}
\end{figure}

Another way to analyze the topology of the resulting networks is
by studying the community structure using an algorithm proposed in
\cite{mod1,mod2}. The algorithm allows us to evaluate the
community structure in a network,  generating a dendrogram
depicting the partition of a network into smaller entities. At
each step the fusion of two structures is proposed and accepted
only if this fusion lowers a global quantity called modularity. A
high modularity is an evidence of having obtained a good partition
of the network under analysis. However, the final result does not
only depends on the network itself, but also on the algorithm used
to find it. Examples of the resulting dendograms are depicted in
Figures \ref{dendo} a and b. The algorithm is finished when the
maximal value of modularity $M$ is reached. What we observe is
that the degree of local organization is higher when the value of
$q$ corresponds to the one presenting the maximum clusterization,
depicted in Fig. \ref{clust}. Both the visible structure of the
dendogram and the higher value of modularity reflect this fact.

\section{Conclusions}

In this work we have presented a model for evolving networks where
the dynamics of the architecture of the links is related to the
affinity between individuals. This aspect associates the model
with that of the stable marriage though in the present case
individuals are not forming couples but groups of affine agents.

One of the most interesting feature is the evolution of the
clustering coefficient as a function of the disorder of the
initial condition. If we associate the parameter $q$ with a degree
of disorder, we can see that the clustering has a maximum for an
intermediate value of $q$. The clustering is an interesting
characteristic of the network since it is related to local
efficiency in transmission of information. Previous model of
networks only displayed a monotonic behavior of this quantity.

On the other hand it is interesting to stress when the system is
driven either by a collective or individual initiative, the
results are qualitatively the same. Though the values of the
energy $E$ where higher for the cases related to individual
dynamics, the overall behavior of the systems is preserved.

We want to end up  by saying that the present work is only the
first stage towards a model of co evolution of the affinity of
agents and the topology of the underlying network. The feedback
between this two features in a more general model can lead to an
collection of interesting results with social relevance. But first
we wanted to isolate those aspects related to the network topology
as was done in a previous work \cite{kup} where it was  the
topology of the network that remained unchanged and the affinity
among agents evolved.


\begin{thebibliography}{99}


\bibitem{gs}  D. Gale, L.S. Shapley, Amer. Math. Monthly {\bf 69}, 9 (1962).

\bibitem{niew}Th.M. Nieuwenhuizen, Physica A {\bf 252}, 178  (1998).

\bibitem{roth} A. Roth and M. A. Sotomayor, Econometrica {\bf 57}, 559 (1989).

\bibitem{roth1} A. E. Roth, Math. Oper. Res. {\bf 7}, 617 (1982).

\bibitem{roth2} A. E. Roth,  J. of Political Economy, {\bf 92}, 991 (1984).


\bibitem{gale1} G. Demance and D. Gale, Econometrica {\bf 53}, 873
(1985).

\bibitem{df}  L. E. Dubins and D. Freeman, Amer. Math. Monthly {\bf 88}, 485
(1981).

\bibitem{ada} H. Adachi, Econom. Lett. {\bf 68}, 43 (2000).

\bibitem{omer} M.-J. Oméro, M. Dzierzawa, M. Marsili, Y.-C. Zhang, J.
Physique I France {\bf 7} 1723 (1997).

\bibitem{dzie} M. Dzierzawa, M.J. Oméro, Physica A {\bf 287}, 321(2000).

\bibitem{calda} G. Caldarelli and A. Capocci.  Physica A {\bf 300}, 325 (2001).

\bibitem{ale} A. Lage-Castellanos and R. Mulet. Physica A {\bf 364}, 389 (2006).


\bibitem{GusIrv} D. Gusfield, R.W. Irving, The stable marriage problem:
structure and algorithms, MIT Press, Cambridge, MA, 1989.

\bibitem{knuth} D.E. Knuth, Stable marriage and its relation to other
combinatorial problems, CRM Proceedings \& Lecture Notes Vol. 10,
AMS, 1997.


\bibitem{kirk} S. Kirkpatrick, C.D. Gelatt, Jr. , M.P.
Vecchi, Science, {\bf 220}, 671 (1983).

\bibitem{met} N. Metropolis, A. Rosenbluth, M. Rosenbluth, A. Teller, E.
Teller, J. Chem. Phys.,{\bf 21}, 1087 (1953).

\bibitem{effi} V. Latora and M. Marchiori, Phys Rev. E. {\bf 87},
19871 (2001)

\bibitem{mod1} M. E. J. Newman and M. Girvan, Phys. Rev. E {\bf 69}, 026113 (2004).

\bibitem{mod2} M. E. J. Newman, Phys. Rev. E {\bf 69}, 066133 (2004).

\bibitem{kup} M. Kuperman, Phys. Rev. E {\bf 73}, 046139 (2006).



\end{thebibliography}
\end{document}